\documentclass[
aps,
pra,
notitlepage,
citeautoscript, superscriptaddress,
reprint
]{revtex4-1}
\usepackage[dvipsnames,usenames]{xcolor}
\usepackage{graphicx}
\usepackage{dcolumn}
\usepackage{bm}
\usepackage{mathrsfs,natbib}
\usepackage[title]{appendix}
\usepackage{graphicx,epsf,amssymb,amsbsy,amsfonts,amssymb,amsmath}
\usepackage[colorlinks=true]{hyperref}
\hypersetup{
    unicode=false,          % non-Latin characters
    pdftoolbar=true,        % show Acrobat
    pdfmenubar=true,        % show Acrobat
    pdffitwindow=false,     % window fit to page when opened
    pdfstartview={FitH},    % fits the width of the page to the window
    pdftitle={},    % title
    pdfauthor={},     % author
    pdfsubject={},   % subject of the document
    pdfcreator={},   % creator of the document
    pdfproducer={}, % producer of the document
    pdfkeywords={} {} {}, % list of keywords
    pdfnewwindow=true,      % links in new window
    colorlinks=true,       % false: boxed links; true: colored links
    linkcolor=magenta, %red,          % color of internal links (change box color with linkbordercolor)
    citecolor=blue,        % color of links to bibliography
    filecolor=magenta,      % color of file links
    urlcolor=blue           % color of external links
}
\usepackage{slashed}
\usepackage{comment}
\usepackage[dvipsnames]{xcolor}
\usepackage{color,soul}
\usepackage{simplewick}

\newcommand{\be}{\begin{equation}}
\newcommand{\ee}{\end{equation}}
\newcommand\beq{\begin{eqnarray}}
\newcommand\eeq{\end{eqnarray}}

\newcommand{\eff}{{\text{eff}}}

\newcommand{\mybar}[1]

\begin{document}

\title{Simulating lattice fermion doubling with a Floquet drive}
\author{Ra\'ul A.\ Brice\~no}%
\email{rbriceno@berkeley.edu}
\affiliation{Department of Physics, University of California, Berkeley, CA 94720, USA}
\affiliation{Nuclear Science Division, 
Lawrence Berkeley National Laboratory, Berkeley, 
CA 94720, USA}
\author{William Gyory}%
\email{wgyory@iastate.edu}
\affiliation{Department of Physics and Astronomy, Iowa State University, Ames, Iowa 50011, USA}%
\author{Thomas Iadecola}
\email{iadecola@iastate.edu}
\affiliation{Department of Physics and Astronomy, Iowa State University, Ames, Iowa 50011, USA}%
\affiliation{Ames National Laboratory, Ames, Iowa 50011, USA}%
\author{Srimoyee Sen}%
\email{srimoyee08@gmail.com}
\affiliation{Department of Physics and Astronomy, Iowa State University, Ames, Iowa 50011, USA}%

\date{\today}

\begin{abstract}
We consider a recently discovered mathematical correspondence between the spectra of a naively discretized lattice fermion and that of a periodically driven (i.e.,~Floquet) quantum system and enhance it into an infrared equivalence between the two systems. 
The equivalence can be framed as a duality relation, allowing us to simulate a two-flavor discrete-time fermion theory on the lattice side, where the two flavors arise from time discretization, using a single-flavor fermion theory on the Floquet side. 
Our demonstration establishes an equivalence between (i) the fermion content, (ii) the correlation functions, and consequently (iii) observables of the two theories in the infrared, going substantially beyond the previously discovered spectral equivalence. 
We also show how interactions may be incorporated into this enhanced infrared equivalence. 
\end{abstract}

\maketitle

The advent of quantum computers offers a novel lens to search for new physics by providing unique ways to constrain predictions from the Standard Model non-perturbatively. 
This has resulted in numerous proposals of new physical observables that are currently inaccessible to classical computing techniques, including time-dependent observables, scattering processes, many-body dynamics of nuclear and neutrino systems, and properties of nuclear matter at finite chemical potential~\cite{Farrell:2024mgu, Turro:2024pxu, Jordan:2012xnu, Marshall:2015mna,Briceno:2020rar, Briceno:2023xcm, Carrillo:2024chu, Davoudi:2024wyv,Jha:2024jan, DiMeglio:2023nsa,Bauer:2022hpo}.
A commonality across these observables is that they are most readily accessible using the Hamiltonian formulation of quantum field theories (QFTs). Despite numerous proposals for Hamiltonian formulations of strongly interacting QFTs, most remain at an early stage of development. Consequently, it is unclear which approach is best suited for studying theories such as quantum chromodynamics, the fundamental theory of the strong nuclear force. This motivates the exploration of novel frameworks that could enhance discovery potential. In this spirit, we present a new formulation for simulating QFTs using Floquet drives.

Digital quantum simulation of a time-independent QFT Hamiltonian $H_{\text{target}}=\sum_i n^i H^i$, with $n^i>0$ and $\sum_i n^i =1$, can be enacted by the unitary circuit $\prod_i e^{-i H^i t^i}$, provided the $t^i$ are sufficiently small.
When the $t^i$ are not necessarily small, this circuit implements evolution under a time-dependent piecewise-constant Hamiltonian, i.e., a periodically driven quantum system with drive period $\sum_i t^i=T$.
Such systems, known as Floquet systems \cite{CayssolReview, RudnerReview, SachaReview, ElseTC, MonroePDTC, KhemaniReview, MonroeDTC, LukinDTC, GoogleDTC, DumitrescuEDSPT, DengFSPT, McIver19, vonKeyserlingk, RudnerReview, Gritsev:2017zdm}, can host exotic non-equilibrium phenomena, including phases of matter that are inaccessible with static Hamiltonians.
Many of these phenomena originate from the fact that energy is no longer conserved in Floquet systems.
Instead, energy is replaced by quasienergy, a periodic variable in frequency space that is only conserved modulo $2\pi/T$ \cite{vonKeyserlingk, RudnerReview, Gritsev:2017zdm}.

Such periodicity in frequency space also arises in action formulations of discrete-time lattice QFTs~\cite{Kanwar:2021tkd, Aarts:1998td, Mou:2013kca, Nielsen:1980rz, Karsten:1981gd}, leading to certain properties that echo the exotic features of Floquet systems. 
Inspired by this apparent similarity, an explicit mathematical equivalence was derived in previous work~\cite{Iadecola:2023uti, Iadecola:2023ekk} between the quasienergy spectrum of a certain Floquet system and that of a discrete-time spectrum of a static Dirac theory in $1+1$ space-time. This spectral equivalence, however, does not necessarily imply that the two theories are equivalent at the level of particle content or correlation functions.

In this work, we demonstrate that the two systems are indeed equivalent: they include the same particle content and share an infrared (IR) correlator equivalence. In this sense, the two can be called IR duals of each other. As we will show, the extent of the correspondence uncovered in this paper suggests that one system can be used to simulate the other. For some other approaches to simulating lattice field theories using Floquet engineering, see \cite{PhysRevD.110.114503,PhysRevResearch.6.L022059, Schweizer_2019, Hayata:2024fnh, Ciavarella:2022tvc}. Although we restrict our attention to a 1+1D system, it may be possible to generalize these ideas to $2+1$D systems, along the lines of Ref.~\cite{Iadecola:2024too}. 

To understand the question of particle content, consider the discrete-time Schr\"odinger equation in $1+1$D,
\beq
i \nabla_{t,t'} \psi = H \psi,
\label{eq:Schrod_v1}
\eeq
where $\nabla_{t,t'}=\frac{\delta_{t,t'-1}-\delta_{t,t'+1}}{2\tau}$ is the naively discretized time derivative with time lattice spacing $\tau$. In frequency space this becomes
\beq
\sin(\omega\tau)\psi=\left(H\tau\right)\psi.
\label{eq:Schrod_v2}
\eeq
For convenience, 
we can choose $H$ to be a lattice Dirac Hamiltonian $H_D$, in which case Eqs.~\eqref{eq:Schrod_v1} and \eqref{eq:Schrod_v2} are Dirac equations.
Denoting the eigenvalues of $H_D$ as $\epsilon_D$, the corresponding eigenstates can have two possible time evolutions: one at frequency $\omega = \tau^{-1} \sin^{-1} \left( \epsilon_D \tau \right)$, and the other at $\omega = \frac{\pi}{\tau} -\tau^{-1} \sin^{-1} \left( \epsilon_D \tau \right)$. 
% This is a manifestation of the well-known fermion doubling in lattice field theory. 
The corresponding lattice Dirac propagator, $\frac{1}{ \gamma^0 \tau^{-1} \sin(\omega \tau) - \gamma^0 H_D},$ features two poles for every $\epsilon_D$ corresponding to these two ``$\pi$-paired" frequencies. 
The two poles arising out of time discretization are naturally interpreted as two flavors of fermions in lattice field theory, a phenomenon known as fermion doubling.
In Floquet systems, $\pi$-pairing can arise for symmetry reasons in certain models, such as that of Ref.~\cite{Iadecola:2023uti}, and this pairing can occur at the level of topologically protected localized edge modes with quasienergies zero and $\pi/T$. Ref.~\cite{Iadecola:2023uti} demonstrated that the superficial similarity between these spectral features extends to an exact spectral mapping.
This work elevates that spectral equivalence to an IR duality.

\begin{figure*}[t]
    \centering
    \includegraphics[width=0.95\textwidth]{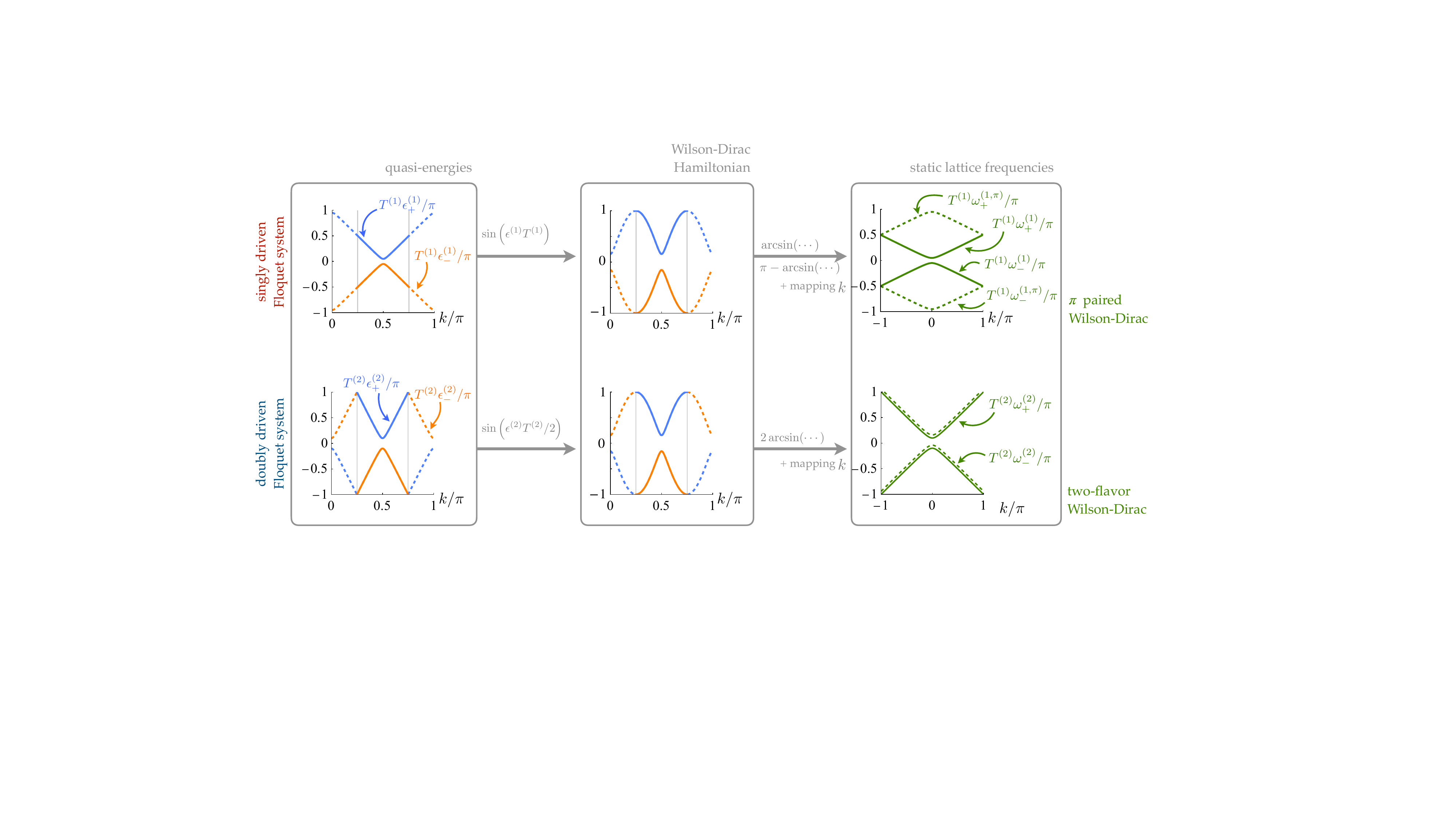}
    \caption{
    In the leftmost panel, we plot the quasienergy eigenvalues with respect to crystal momenta $k$ for the singly driven (top) and doubly driven (bottom) cases. The center panel plots the eigenvalues of the target Hamiltonian on the discrete-time side. The Hamiltonian and eigenvalues of the two cases are identical. The right panel shows the solutions to the discrete-time equation of motion that reproduce the Floquet quasienergy eigenvalues from the left panel, again in both the singly driven (top) and doubly driven (bottom) cases.
    The plots correspond to parameter values $t_0 = \frac\pi4$, $t_1 = (1.1) \frac\pi4$, i.e., we have picked a gapped region of the spectra to illustrate the equivalence.
    }
    \label{eff2}
\end{figure*}

\textbf{\textit{Floquet model.}}---The properties of Floquet systems at ``stroboscopic times" (i.e., integer multiples of $T$) are described in terms of the evolution operator $U(T)$ via the stroboscopic Hamiltonian
\beq
H_S=\frac{i}{T}\log(U(T)).
\eeq
Its eigenvalues, the quasienergies, are conserved modulo $2\pi/T$ and defined within a Brillouin zone (BZ) from $-\pi/T$ to $\pi/T$.
We begin with the model of Ref.~\cite{Iadecola:2023uti}, defined on a lattice of $2N$ sites in one spatial dimension, with evolution operator
\begin{align}
U^{(1)}(t)&=\begin{cases}e^{-i H_0 t} & \text{for } 0<t<t_0\\
e^{-i H_1(t-t_0)}e^{-i H_0 t_0} & \text{for } t_0\leq t<t_0+t_1
\end{cases},
\label{eq:U}
\end{align}
with
\begin{align}
\begin{split}
H_0&=2\sum^{N-1}_{j=0}(a_{2j}^{\dagger}a_{2j+1}+\text{H.c.})\\
H_1&=2\sum^{N-1}_{j=0}
(a_{2j+1}^{\dagger}a_{2j+2}+\text{H.c.})
\label{h}
\end{split}
\end{align}
and $t_0+t_1=T^{(1)}$. 
Here $a_i^{(\dagger)}$ denotes a complex fermion annihilation (creation) operator at site $i = 0, 1, \dots, 2N - 1.$ 
$H_0$ and $H_1$ are essentially the two terms in the Hamiltonian of the Su–Schrieffer–Heeger (SSH) model \cite{SSH}, which itself describes a lattice Dirac fermion in one spatial dimension. 
In this work, we restrict our attention to systems with spatial periodic boundary conditions (PBC). 

To construct a two-flavor theory,
we will modify the drive in Eq.~\eqref{eq:U} to 
\begin{align}
U^{(2)}(t)&=\begin{cases}U(t) & 0<t<T^{(1)}\\
U(t-T^{(1)})U(T^{(1)}) & T^{(1)}\leq t<2T^{(1)}\\
\end{cases}
\label{eq:U2}
\end{align}
and restrict to $t_0=\pi/4$ for reasons that we will explain later in the text. 
Since we are interested in stroboscopic dynamics at $2nT^{(1)}$ with $n\in \bf{Z}$,
the corresponding effective Hamiltonian for this drive is given by $H^{(2)},$ where 
\beq
e^{-i H_1 t_1}e^{-i H_0 t_0}e^{-i H_1 t_1}e^{-i H_0 t_0}\equiv e^{-i H^{(2)} T^{(2)}},
\label{eff}
\eeq
where $T^{(2)}=2T^{(1)}$.
We will refer to $U^{(1)}$ and $U^{(2)}$ as the singly and doubly driven Floquet models, respectively.

Following Ref.~\cite{Iadecola:2023uti}, the eigenvalues of $U^{(1)}(T^{(1)})$ are
\begin{align}
\lambda_\pm(k)= Q(k) \mp i \sqrt{1 - Q^2(k)},
\end{align}
where $k = 0, \frac\pi N, \frac{2\pi}N,\dots,\frac{(N-1)\pi}N$ is the spatial momentum, and $Q(k)$ is defined as 
\begin{align}
    \label{Q}
    Q(k) &= \cos(2t_0)\cos(2t_1)
    - \cos(2k)
    \sin(2t_0)\sin(2t_1).
\end{align}
Given the eigenvalues of $U^{(1)}(T^{(1)})$, the corresponding quasienergies are
\begin{align}
    \label{eq:eps_single}
    \epsilon_{\pm}^{(1)}(k)\,\, T^{(1)}
    &= i
    \log(\lambda_\pm(k)).
\end{align}
From this expression, it is evident that for  $t_0=\pi/4$, the spectrum has a special feature: it is $\pi$-paired in the sense that for $\frac\pi4 \leq k < \frac{3\pi}4$ we have $\epsilon^{(1)}(k + \pi/2) = \frac\pi{T^{(1)}} - \epsilon^{(1)}(k)$. 
To make this further evident, we plot $\epsilon_{\pm}^{(1)}$ as a function of $k$ in the top-left panel of Fig.~\ref{eff2} for $t_0=\frac{\pi}{4}$ and $t_1=(1.1)\frac{\pi}{4}$. The solid and dashed bands carry eigenvalues that are $\pi$-pairs of each other, while no two eigenvalues within each band are $\pi$-pairs. 
The eigenvalues of $U^{(2)}(T^{(2)})$ are simply $\lambda_\pm^2(k)$, corresponding to quasienergies
\begin{align}
    \label{eq:eps_double}
    \epsilon^{(2)}_{\pm}(k) T^{(2)}
    &=2\epsilon_{\pm}^{(1)}T^{(1)} \,\,\,\,\text{mod} \,\,\,2\pi,
\end{align}
where we use the convention $x \text{  mod } 2\pi \in [-\pi, \pi)$.  
These quasienergy eigenvalues are plotted in the bottom-left panel of Fig.~\ref{eff2} for $t_0=\frac{\pi}{4}$ and $t_1=(1.1)\frac{\pi}{4}$. 
Doubling the drive has brought ``dashed" modes in the top-left panel, which were previously $\pi$-paired with the ``solid" modes, into degeneracy with ``solid" modes in the opposite band in the bottom-left panel. We can reinterpret this degeneracy as two fermion flavors, and we now attempt to make a concrete connection between this two-flavor theory and a discrete-time static theory.

For convenience, we define $\epsilon^{(1/2)}(k) \equiv |\epsilon_{+}^{(1/2)}(k)|$. In what follows, we will at times only refer to $\epsilon^{(1/2)}(k)$ in our formulae, since analogous claims will hold for ${-}\epsilon^{(1/2)}(k)$.

\textbf{\textit{Two-flavor Wilson-Dirac theory.}}---We now turn our attention to the construction of the time-discretized Wilson-Dirac theory that can be related to the Floquet models discussed above. 
We will start with the singly driven system, the correspondence for which was derived in \cite{Iadecola:2023uti} and is reviewed here. The correspondence for the doubly driven system, which is described in the subsequent text, is the new result that gives rise to an equivalence/duality. 
To construct the target Hamiltonian of the static discrete time theory for the singly driven case, we are guided by Eq.~\eqref{eq:Schrod_v2} with time lattice spacing $\tau$ set to $T^{(1)}$. 
For the lattice spectrum of $H$ to match the quasienergy eigenvalues $\epsilon^{(1)}$, it must have eigenvalues $\sin(\epsilon^{(1)}T^{(1)})$, with momenta restricted to half the possible values to avoid double counting.
Thus we choose to restrict to momenta corresponding to the ``solid" bands in Fig.~\ref{eff2}, resulting in the equation of motion 
\begin{align}
    \sin\left(\omega \,\,T^{(1)}\right)=\sin \left(\epsilon^{(1)} \left(\frac{k}{4}+\frac{\pi}{2}\right)T^{(1)}\right),
    \label{e4}
\end{align} 
where $-\pi\leq k<\pi$, from which it immediately follows that the solutions are $T^{(1)}\omega^{(1)}=T^{(1)}\epsilon^{(1)}$ and $\pi-T^{(1)}\epsilon^{(1)}$
as illustrated in the top-right panel of Fig.~\ref{eff2}. 

Ref.~\cite{Iadecola:2023uti} finds that the RHS of Eq.~\eqref{e4} equals $T^{(1)}\kappa(k)$ where $\kappa(k)$ are the eigenvalues of a Wilson-Dirac Hamiltonian of the form
\beq
H_{\rm WD} = R\, \gamma^0 \gamma^1 \sin k + \gamma^0\, [m + R(1 - \cos k)],
\label{eq:HWD}
\eeq
with $\gamma^0=\sigma_1, \gamma^1=i\sigma_2$, $R = \frac1{2T^{(1)}} - \frac m2$, $m=\frac{\sin 2\eta}{T^{(1)}}$, $\eta=t_1-\frac{\pi}{4}$ and
\beq
\kappa(k) = \sqrt{\left(R \sin k \right)^2 + [m + R(1 - \cos k)]^2}.
\eeq
Thus we see that the discrete time theory with Hamiltonian $H_{\rm WD}$ reproduces the top figure of the third panel of Fig.~\ref{eff2}, establishing its spectral equivalence with the singly driven Floquet model.

To obtain an analogous equivalence for the doubly driven system, we rewrite Eq.~\eqref{e4} by replacing $T^{(1)}$ by $T^{(2)}/2$ as 
\beq
\sin (\omega T^{(2)}/2)= \kappa T^{(2)}/2.
\label{eql1}
\eeq 
The frequency solutions for this equation are $\omega^{(2)} T^{(2)} = 2\sin^{-1}(\kappa T^{(1)})$ and $2\pi - 2\sin^{-1}(\kappa T^{(1)})$,
the second of which must be dropped as it is outside the BZ.
We are left with $\omega^{(2)} T^{(2)}=\epsilon^{(2)} T^{(2)}$, 
which can be seen using Eq.~\eqref{eq:eps_double}.

We now address what kind of a discrete-time lattice action will result in an equation of motion of the form Eq.~\eqref{eql1}. 
Note that a naively discretized time derivative with lattice spacing $T^{(2)}/2$, in a Schr\"odinger equation with lattice Hamiltonian eigenvalues $\kappa$, will produce Eq.~\eqref{eql1}, but with a frequency BZ that extends from $-\pi/(T^{(2)}/2)$ to $\pi/(T^{(2)}/2)$, which while appropriate for the singly driven case is undesirable for the doubly driven case. 
For the latter, we want Eq.~\eqref{eql1} to hold with a BZ between $-\pi/T^{(2)}$ and $\pi/T^{(2)}$.

Inspired by staggered fermions \cite{Kronfeld:2007ek, PhysRevD.13.1043}, we find a perfectly local lattice action that can give rise to Eq.~\eqref{eql1} while maintaining the desired BZ. 
The time lattice in this case is divided into two sublattices, each unit cell containing two lattice sites, with unit cell separation $T^{(2)}$ and intra-cell lattice spacing $T^{(2)}/2$. 
The action is given by
\beq
S=\int \frac{d^2k}{(2\pi)^2} \bar\psi[\mathbf{P}^0\otimes \gamma^0-1_{2\times 2}\otimes (\gamma^0 H_{\rm WD})]\psi,
\label{eq:S_v0_1}
\eeq
where $d^2k = d\omega \, dk$ and we have introduced the operator
\beq
\mathbf{P}^0=\frac{1}{T^{(2)}}\begin{pmatrix}
0 && 1-e^{-i \omega T^{(2)}}\\
1-e^{i \omega T^{(2)}} && 0
\end{pmatrix},
\label{P0eq}
\eeq
where the two-by-two subspace of $P_0$ corresponds to the sublattice space and $\bar\psi=\psi^{\dagger}(1_{2\times 2}\otimes\gamma^0)$. 
 
It is straightforward to see that the eigenvalues of $\mathbf{P}^0$ are $\pm\frac{\sin(\omega T^{(2)} /2)}{T^{(2)}/2}$, which results in the equation of motion with spectra given by Eq.~\eqref{eql1}.
The operator $\mathbf{P}^0$ can be expressed as the Fourier transform of a local operator defined in discrete time,
\beq
\widetilde{\mathbf{P}}^0=(-1)^{i+1}\frac{\left(\delta_{i,j-1}-\delta_{i,j+1}\right)}{T^{(2)}},
\eeq
where $i, j$ denote lattice sites in time; this justifies the previous claim about locality of the action. Interestingly, the presence of two sublattices implies double degeneracy for each frequency solution to Eq.~\eqref{eql1} for a specific momentum, which leads to the interpretation of the two sublattice indices as two flavors, a manifestation of fermion doubling. The spectrum is illustrated in the bottom-right panel of Fig.~\ref{eff2}.
The equivalence between lattice and doubly driven Floquet spectra is then simple: we identify the two sublattice indices (the two flavors) on the discrete-time side with the two Floquet bands between $-\frac{\pi}{4}\leq k <\frac{\pi}{4}$ and $\frac{\pi}{4}\leq k <\frac{3\pi}{4}$ in the bottom-left panel of Fig.~\ref{eff2}.

\textbf{\textit{Correlator map.}}---Having demonstrated a spectral equivalence between the Floquet and lattice theories, we now move on to correlation functions.
This reduces to showing an equivalence between the single-particle propagators of the two theories.
We begin with the lattice side.
We implement a change of basis, $\tilde{\psi}=F'\psi,$ such that $\mathbf{P}^0$ is diagonalized (see Sec.~A of the supplemental materials).
We now slightly modify our notation, explicitly writing $\tilde\psi_{a, i},$ where $a=1,2$ is the time sublattice index and $i$ is the spin/Dirac/Pauli index.  
Furthermore, defining $\tilde{\chi}_{a}(\omega)=\tilde{\psi}_{a}((-1)^{a-1} \omega),$
the action can be expressed as
\beq
S_2=\sum_a\int_{-\pi}^{\pi} \frac{dk}{2\pi}\int_{-\frac{\pi}{T^{(2)}}}^{\frac{\pi}{T^{(2)}}} \frac{d\omega}{2\pi} \tilde{\chi}_a^{\dagger}\left[\frac{\sin \frac{\omega T^{(2)}}{2}}{T^{(2)}/2}-H_{\text{WD}}\right]\tilde{\chi}_a\nonumber.
\label{min3}
\eeq
The propagator for $\tilde{\chi}$ is then given by
\begin{align}
\tilde G(\omega, k)_{ab} 
&= \left(\frac{\frac{\sin \frac{\omega T^{(2)}}{2}}{T^{(2)}/2}+H_{\text{WD}}\,}{\frac{\sin^2\frac{\omega T^{(2)}}{2}}{(T^{(2)}/2)^2}-\kappa^2}\right)\gamma^0\delta_{ab}\equiv \tilde{G}^{a}\delta_{ab}.
\end{align}

We now turn to the Floquet side. 
The stroboscopic dynamics are described by the action 
\beq
S_\text{str}=\int \frac{d^2k}{(2\pi^2)}\phi^{\dagger}(\omega-H^{(2)})\phi,
\label{S_str}
\eeq
where the Floquet Hamiltonian for the doubly driven system, $H^{(2)}$, is treated as a static Hamiltonian, the spatial $k$ integral goes from $0$ to $\pi$, and the $\omega$ integral goes from $-\infty$ to $\infty$. 
We can again change basis, $J\tilde\phi = \phi$ (see Sec.~A of the supplemental materials), so that the Floquet stroboscopic Hamiltonian becomes $H_{\text{final}} = \epsilon_0 \sigma_1 - \sqrt{(\epsilon^{(2)}(k))^2 - \epsilon_0^2} \,\, \sigma_3 = \gamma^0 \epsilon_0 + \gamma^0 \gamma^1 \sqrt{(\epsilon^{(2)}(k))^2 - \epsilon_0^2},$ where we have defined $\epsilon_0 \equiv |\epsilon^{(2)}(k=0)|$, i.e., the magnitude of the gap. The Floquet stroboscopic Hamiltonian is now expressed in terms of a gamma matrix basis analogous to the Wilson-Dirac Hamiltonian $H_\text{WD}$. 
The propagator for $\tilde{\phi}$ is given by
\beq
G_{\text{str}}(\omega, k) 
=  \left(\frac{\omega + H_{\text{final}}}{{\omega}^2 - {(\epsilon^{(2)}(k))^2}}\right)\gamma^{0}.
\label{e2}
\eeq

To compare $G_{\text{str}}$ with the lattice propagator, we can identify two flavors in $G_{\text{str}}$ by defining:
\beq
G_{\text{str}}^1(\omega,k)&\equiv& G_{\text{str}}\left(\omega,\frac{k+2\pi}{4}\right)\nonumber\\
G_{\text{str}}^2(\omega,k)&\equiv& G_{\text{str}}\left(\omega,\frac{k}{4}\right)\nonumber\\
G_{\text{str},ab}^F&\equiv &G_{\text{str}}^{a}\,\,\delta_{ab},
\eeq
where $-\pi\leq k<\pi$ and $a, b$ take values $1, 2,$ indicating two flavors.
The explicit correspondence can be seen by comparing $\tilde G$ to $G_\text{str}^F$. 
It is clear that the pole positions for each flavor on the two sides of the correspondence match, and the propagators are identical in the small $\omega, k$ limit (hereafter referred to as the IR limit) and small $\eta$ expansion at leading order. Note that, since the flavors are degenerate, we could permute the flavors on one side of the correspondence and the results would still hold. It is also instructive to compute lattice correlators in time and compare with the corresponding quantity on the Floquet side. We do this in Sec.~B of the supplemental materials.

\textbf{\textit{Toward interactions.}}---Given the correspondence between correlators, one may construct a map relating possible perturbative interactions between the two sides. 
For example, consider the current-current (Thirring) interaction on the lattice side:
\beq
\lambda\int d^2x \left(\bar{\tilde{\chi}}_a\gamma^\mu \tilde{\chi}_a \right)\left(\bar{\tilde{\chi}}_a\gamma_\mu \tilde{\chi}_a \right)
\label{latint}
\eeq
where $\tilde{\chi}(t,x)$ here stands for the Fourier transform of $\tilde{\chi}(\omega, k)$.
To obtain an interaction on the Floquet side that corresponds to this one in the IR limit, one could deform the Floquet Hamiltonian with an analogous piece 
\beq
H_{\text{int}}
=\lambda\int dx\left(\bar{\tilde{\phi}}\gamma^\mu \tilde{\phi}\right)\left(\bar{\tilde{\phi}}\gamma_\mu \tilde{\phi} \right)
\eeq
where $\tilde{\phi}(t,x)$ here is the Fourier transform of the field $\tilde{\phi}(\omega, k)$.
The IR limit is essentially a low-momentum requirement, i.e., the limit $|k| \ll \pi$ on the lattice side and $|k|\ll\pi/4$ and $|k-\pi/2|\ll \pi/4$ on the Floquet side. We will refer to all these limits as low momentum conditions. 
This condition arises out of the observation that the form of the lattice interaction proposed here conserves momentum. 
It also conserves flavor, and the interaction only involves fields of the same flavor. 
On the Floquet side, we have identified the regions in momenta from $-\pi/4\leq k<\pi/4$ and $\pi/4\leq k<3\pi/4$ as two different flavors. 
Therefore, in order to correctly replicate this local lattice interaction, we must find a local interaction on the Floquet side with the following property: Each vertex should not only conserve momentum, but should also only allow incoming and outgoing states that carry momenta within one of the two branches, either $-\pi/4\leq k<\pi/4$ or $\pi/4\leq k<3\pi/4$.

It is not easy to find a simple driving protocol inducing such a deformation. 
For example, consider the drive protocol 
\begin{align}
\tilde{U} 
&= \begin{cases}
U_2(t) \hspace{1in} & 0<t<T^{(2)}\\
U_\delta U_2(T^{(2)}) & T^{(2)}<t<T^{(2)}+\delta T,\\
\end{cases} \\
U_\delta &= e^{-i g\,\delta T\int dx \left(\bar{\tilde{\phi}}\gamma^\mu \tilde{\phi} \right)\left(\bar{\tilde{\phi}}\gamma_\mu \tilde{\phi} \right)}
\label{drive}
\end{align}
where $g\delta T=\lambda T^{(2)}$ and the stroboscopic time is $T^{(2)} \allowbreak + \allowbreak\delta T  \allowbreak = \tilde{T}\approx T^{(2)},$ assuming $(\delta T)/T^{(2)} \ll 1$. The Baker-Campbell-Hausdorff (BCH) formula gives $\tilde{U}(\tilde T)=e^{-i H_\eff T^{(2)} -i H_\text{int}\delta T-\frac{1}{2}[H_{\text{int}}\delta T, H_\eff T^{(2)}]+\cdots }$. 
It is then clear that this driving protocol generates the desired interaction $H_\text{int}$, along with many nonlocal terms, including some that are linear in $\delta T$, arising out of the nested commutators. 
In general, there is no way to remove these nonlocal terms with the driving protocol we have adopted here. 
However, with this class of protocol, if one is interested in the long-wavelength limit of certain scattering observables at leading order in $\lambda$, e.g., scattering matrix elements for a two-to-two scattering process, the contributions from the commutators are suppressed, enabling a correspondence between the weakly interacting lattice and Floquet systems. 
We discuss the details in Sec.~C of the supplemental materials. 
The correspondence may go beyond the weak coupling and long wavelength limits. 
However, such a discussion is beyond the scope of this paper. Furthermore, the choice of current-current interaction in the above analysis was made merely for the purpose of illustration. 
It should be possible to create similar protocols for other types of interactions, including gauge interactions. 
For introducing gauge interactions in particular, it is important to remember the exact symmetries of the system, since those are the most natural candidates for gauging. In this specific example, one could consider gauging the exact global $U(1)$ symmetry of the theory.
We leave these topics open for future investigations, which have been made possible due to the equivalence developed here.

\textbf{\textit{Conclusion.}}---Previous work demonstrated that free, static discrete-time theories and periodically driven (Floquet) free systems in continuous time can share equivalent spectra. 
Naive time discretization in the former leads to fermion doubling, i.e., the emergence of a two-flavor theory due to discretization. 
This naturally raises the question: {\it Can the Floquet spectrum of the driven system be reinterpreted as that of a two-flavor theory?} 
In this work, we have demonstrated that this is indeed the case.
This observation can be further exploited to engineer an IR mapping between correlation functions and observables in Floquet and lattice systems, implying a duality between the two. 
Furthermore, we have shown through examples how one may extend the IR equivalence of the free theories to theories with perturbative interactions. 
Adding interactions generically leads to a Floquet Hamiltonian containing infinitely many BCH commutators between the free Hamiltonian and the interaction piece.
The effect of these commutators, however, can be suppressed for certain scattering processes in the IR limit, leading to an IR equivalence between interacting Floquet and discrete-time theories. 
Whether more efficient protocols can be found that eliminate the undesired BCH terms altogether should be explored in future work. Floquet systems have also been used to engineer free unpaired Weyl fermions on the lattice \cite{PhysRevLett.123.066403}. With possible application to chiral gauge theories in mind, our proposals for introducing interactions could be extended to such systems.
It is also worth exploring whether the equivalence survives under strong interactions.
Since Floquet systems are naturally realized on quantum computers, this work suggests a new path toward quantum simulation of lattice field theories.

\section*{Acknowledgments}

RAB was supported in part by the U.S. Department of Energy, Office of Science, Office of Nuclear Physics under Awards No.~DE-AC02-05CH11231. 
S.S.\ and W.G.\ acknowledge support from the U.S.\ Department of Energy, Nuclear Physics Quantum Horizons program through the Early Career Award DE-SC0021892.
T.I.\ acknowledges support from the National Science Foundation under Grant No.~DMR-2143635.

\bibliography{floquet1.bib}

\end{document}